\documentclass[referee]{raa}  
\usepackage{natbib}
\usepackage{graphicx,times}             
\usepackage{amssymb,amsmath,mathrsfs}
\usepackage{newtxtext,newtxmath}
\bibpunct{(}{)}{;}{a}{}{,}
\usepackage[pagebackref=true]{hyperref}
\usepackage[main=english]{babel}
\usepackage{babel}

\newcommand{\bea}{\begin{eqnarray}}
\newcommand{\eea}{\end{eqnarray}}

\begin{document}

\title{fiDrizzle-MU: A Fast Iterative Drizzle with Multiplicative Updates}

 \volnopage{ {\bf 2025} Vol., {\bf X} No. {\bf XX}, 000--000}
 \setcounter{page}{1}
 
\author{Shen Zhang\inst{1,2,3}, Lei Wang\inst{4,5,6}\thanks{E-mail: leiwang@pmo.ac.cn}, Huanyuan Shan\inst{1,3,7}, Ran Li\inst{1,2,3}, Xiaoyue Cao\inst{1,2,3}, Yunhao Gao\inst{1,3}}
  \institute{
   $^1$ National Astronomical Observatories, CAS, A20 Datun Road, Chaoyang District, Beijing, 100101, China;\\
   $^2$ School of Physics and Astronomy, Beijing Normal University,  Beijing 100875, China;\\
   $^3$ School of Astronomy and Space Science, University of Chinese Academy of Sciences, Beijing 100049, China;\\
   $^4$ Purple Mountain Observatory, Chinese Academy of Sciences, No. 10 Yuan Hua Road, Nanjing 210023, China;\\
   $^5$ National Basic Science Data Center, Building No.2, 4, Zhongguancun South 4th Street, Haidian District, Beijing 190, China;\\
   $^6$ Zhejiang University-Purple Mountain Observatory Joint Research Center for Astronomy, Zhejiang University, Hangzhou 327, China;\\
   $^7$ Shanghai Astronomical Observatory, CAS, Nandan Road 80, Shanghai 200030, China
  } 
   
\vs \no
 {\small Received 2025 January XX; accepted XXXX February XX}



\abstract{
In this paper, we introduce a new algorithm, {\it fiDrizzle-MU}, to coadd multiple exposures with multiplicative updates in each iteration instead of the difference correction terms of the preceding version. We find multiplicative update mechanisms demonstrate superior performance in decorrelating adjacent pixels compared to additive approaches, reducing noise complexity in the final stacked images. After applying {\it fiDrizzle-MU} to the JWST-NIRCam F277W band data, we obtain a comprehensive reconstruction of a potential gravitationally lensed quasar candidate substantially blurred by the JWST pipeline's resampling process.
\keywords{Methods: analytical -- Techniques: image processing -- Gravitational lensing: strong}
}

 \authorrunning{Shen Zhang et al. }    
 \titlerunning{fiDrizzle-MU}  
 \maketitle
\section{Introduction}
\label{sect:intro}

The rapid deployment of astronomical telescopes has been driving the production of vast amounts of observational data, which is posing significant challenges to the data processing methodologies and capabilities within the astronomical community. One critical aspect is the combination of multiple exposures of under-sampled images to attain certain degrees of image qualities (e.g. increasing signal-to-noise ratios, upgrading the image resolution, diminishing noises). In general, astronomical telescopes aim to achieve a large field of view during observations. However, due to a variety of factors, including but not limited to cost-effectiveness, the need to control readout noise, technical limitations, the number of detectors in imaging array is typically insufficient to meet the Nyquist sampling criterion(\citealt{Nyquist}; \citealt{Shannon}), resulting in under-sampling. Many telescopes are undersampled\citep{Fruchter+2002}, at least in some bands. For instance, the pixel size of Hubble Space Telescope - Wide Field Camera (HST-WFC) is $0.046^{\prime \prime}$, while the angular resolution of HST is $0.05^{\prime \prime}$. According to the Nyquist sampling theorem, this represents severe undersampling. Additionally, James Webb Space Telescope - Near Infrared Camera (JWST-NIRCam) is undersampled for short wavelengths below $2 \mu m$ and long wavelengths below $4 \mu m$(\citealt{Makidon+2007}; \citealt{Wang+2025}). In under-sampled images, components with different frequencies are blurred with each other. This phenomenon is referred to as aliasing. Recovering additional details from under-sampled images often requires the use of a dither strategy(\citealt{Lauer+1999}; \citealt{Hook+2000}), which means that small but random shifts between exposures are introduced, thereby reducing aliasing and improving spatial resolution.

\begin{figure}
\centering
\includegraphics[width=6.1in]{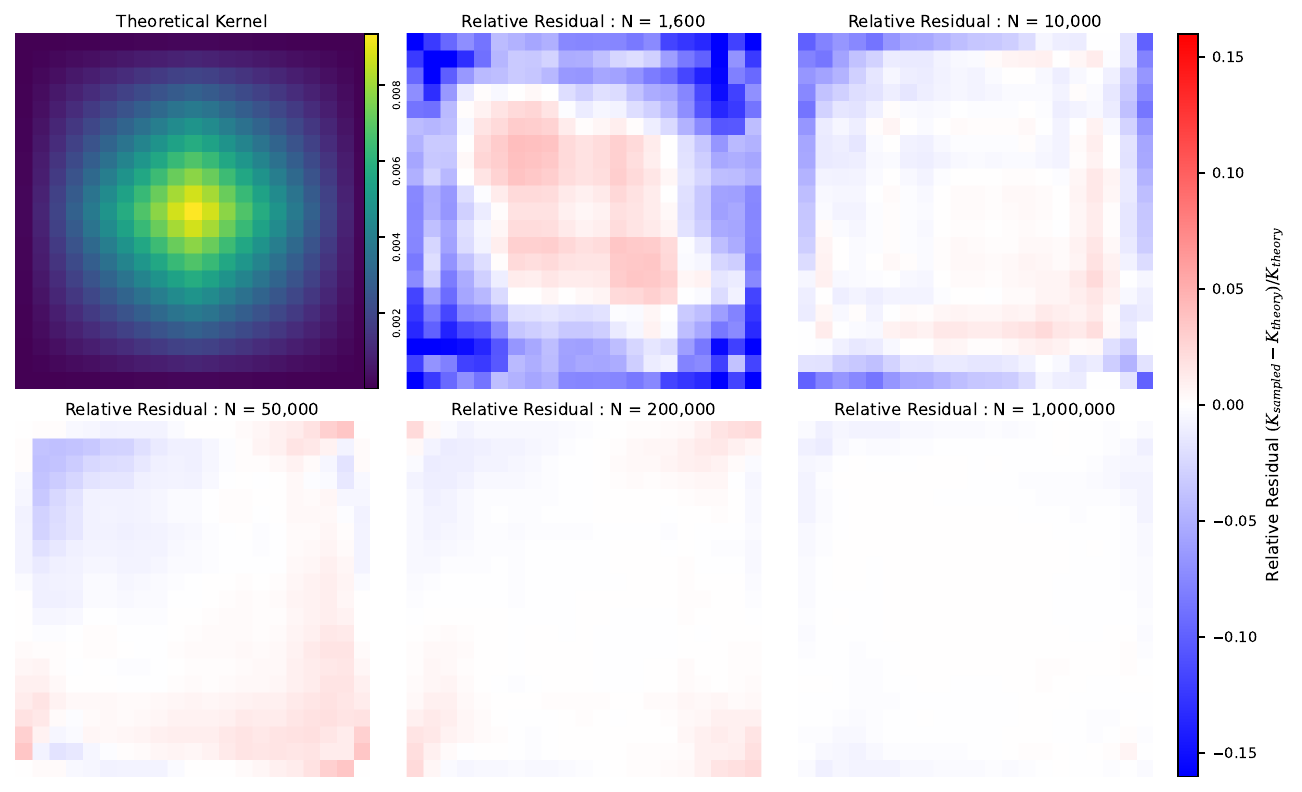} 
\caption{Comparison between the convolution kernel derived from the pixelation blurring simulation and the theoretical prediction. Each convolution kernel in the figure corresponds to a Drizzled image pixel, derived under conditions of a dither pattern consisting solely of sub-pixel uniform random shifts along with a tenfold increase in sampling rate, delineates the flux redistribution process. The original pixel value of a target pixel, intended for restoration, is dispersed across a $21 \times 21$ pixel matrix centered at its location. The upper left panel illustrates the theoretical outcome obtained with an infinite ensemble of nearly continuous dithered frames. The subsequent five panels display the relative residuals between the convolution kernels derived from the Drizzled outcomes and the theoretical kernel. These residuals are evaluated for datasets consisting of 1,600, 10,000, 50,000, 200,000, and 1,000,000 dithered frames, respectively.} \label{fig_kernel}
\end{figure}

For the purpose of coadding multi-exposure images, i.e. the dithered frames, the community have developed several valuable and enlightening methods, such as {\it interlacing}, {\it shift-and-add} (\citealt{Bates+1980, Farsiu+2004b}) and {\it Drizzle} (\citealt{Fruchter+2002}). In this regard, {\it Drizzle} which draws upon advantageous aspects of both {\it interlacing} and {\it shift-and-add}, has become the de facto standard for the combination of images photographed by HST and JWST. Despite its effectiveness in restoring high-resolution images of resolved objects, Drizzle struggles with the reconstruction of unresolved structures and high-frequency details \citep{Fruchter+2011}. These high-frequency components correspond to sharp intensity transitions and small-scale features, which are crucial for characterizing compact sources and preserving morphological fidelity. However, they are particularly vulnerable to information loss due to undersampling, noise and the blurring effects of the instrument optics. Meanwhile, the Drizzle algorithm distributes the flux of a single input pixel across several output pixels, which breaks the independence between output pixels and leads to the emergence of noise correlation. Mathematically, CCD sampling, dithering, and resampling spread output pixel flux into adjacent pixels—analogous to convolving with a diffusion kernel, which can be effectively modeled by the so-called \textit{Note Spike} profile (see the appendix in \citealt{Wang+2022} for reference). The upper left panel of Figure \ref{fig_kernel} depicts the convolution kernel according to this profile with an up-sampling factor of 10. {\it Drizzle} involves two parameters, {\tt pixfrac} ({\it p}) and {\tt PSR} ({\it s}), which represent the linear scaling ratio between the shrunk input pixels and the original input pixels, and that between the output pixels and the original input pixels. As anticipated from an intuitive perspective, a lower {\it p/s} reduces the pixel correlations in the output Drizzled image. The trade-off, however, is the amplification of the high-frequency artifacts, as the noise from every single input pixel is condensed into fewer output pixels, resulting in a lack of smoothing.


{\it iDrizzle} (\citealt{Fruchter+2011}) was proposed as an refined variant of Drizzle. The framework of this algorithm was designed in an iterative fashion. In each iteration, a low pass filter would be used after oversampling, which effectively suppresses high-frequency artifacts. And as the updates carry on, the {\tt pixfrac} decreases gradually, which mitigates noise correlations due to Drizzling. However the oversampling - low pass filtering - interpolating process decelerates the convergence and slows down the computational speed. The application of the tapering function in Fourier domain turns white noises into reddish ones which is another cause of noise correlation. \cite{Wang+2017} developed another iterative {\it Drizzle}-based method {\it fiDrizzle} which has a difference-correction term (hereafter {\it fiDrizzle-DC}). {\it fiDrizzle-DC} proves more computationally efficient compared with {\it iDrizzle}, and performs better than {\it iDrizzle} given the same number of iterations. Without the use of tapering function and operations in Fourier domain, {\it fiDrizzle-DC} suffers less from noise correlation compared with {\it iDrizzle}. However, due to its difference-correction term, {\it fiDrizzle-DC} is relatively sensitive to the ringing effect. Additionally, the difference-based updates still appear to be insufficiently efficient in signal deblending and pixel decorrelation.

 In this work, we modify {\it fiDrizzle} with the application of a multiplicative update in each iteration. The multiplicative updates make the programme converging to the optimal reconstruction with fewer iterations. Apart from this, we impose positivity constraints on all the approximations, which suppress the ringing effect to a significant degree. With all these measures, {\it fiDrizzle-MU} could restore high-resolution images with extensive details from under-sampled dithered frames, resolve sources with minimal spatial separation and invert the Drizzling kernel to concentrate the aliased signals from peripheral electronic pixels while alleviate noise correlation and artificially ringing effect. The paper is organized as follows: We describe the {\it fiDrizzle} algorithm in detail in Section \ref{sect:method}, and illustrate its reconstruction power and converging efficiency in Section \ref{sect:analy} --- the complete characterization of a gravitationally lensed quasar system that had previously been only partially resolved in existing JWST data enabled by the astrometric precision of our algorithm, represents an interesting outcome of this investigation, and de-pixelation power of the algorithm is discussed quantitatively which constitutes the primary focus of this work. In Section \ref{sect:comput}, we evaluate the significant advantage of {\it fiDrizzle-MU} over previous algorithms in terms of computational consumption. We present a comprehensive discussion of the results and provide conclusive remarks in Section \ref{sect:discu&conclu}.

\begin{table}[htbp]
\centering
\caption{Important Notations, Symbols, and Abbreviations Used in the Paper}
\label{tab:notation}
\begin{tabular}{ll}
\hline
\textbf{Symbol/Variable} & \textbf{Definition/Description} \\
\hline
$F_i$ & Image approximation at the $i$th iteration; $F_0$ denotes the initial Drizzled image \\
$I^k$ & The $k-$th dithered exposure image \\
$L_E$ & Effective overlapping layer (normalization factor) \\
$\mathfrak{S}^k_u\left\{\cdot\right\}$ & Upsampling operator for the $k$th exposure \\
$\mathfrak{S}^k_d\left\{\cdot\right\}$ & Downsampling operator for the $k$th exposure \\
$\frac{1}{L_E} \sum_{k=1}^{N} \mathfrak{S}^k_u\left\{\cdot\right\}$ & \textit{Drizzle} operator \\
$\gamma$ & Iterative step-size parameter (set as $\gamma=1$ in this paper) \\
\texttt{pixfrac} & the linear scaling ratio between the shrunk input pixels and the original input pixels \\
\texttt{PSR} & pixel scale ratio, ratio of input to output pixel scale\citep{Fruchter+2002,Makidon+2007} \\
\hline
$L_t$ & True brightness distribution of the source \\
$K_o$ & Telescope point spread function (PSF) \\
$P_f$ & Fine grid sampling operator \\
$P_{e}^k$ & Sampling operator corresponding to the electronic pixels of the $k$th exposure \\
$W^k$ & Weight map for the $k-$th exposure \\
$L_a$ & Approximation to the PSF-blurred brightness distribution --- the \textit{Drizzle} result \\
$K_d$ & Dithering kernel, defined as the weighted sum of sub-dithering kernels \\
$\mathscr{K}_k^d$ & Sub-dithering kernel for the $k$th exposure \\
$\mathscr{K}_k^{d*}$ & Adjoint (transpose) operator of $\mathcal{K}_k^d$ \\
\hline
\multicolumn{2}{l}{\textbf{Reconstruction Quality Metrics:}} \\
PSNR & Peak signal-to-noise ratio, measuring the fidelity of a reconstructed image \\
OCFR & Off-Center Flux Ratio, representing the proportion of flux outside the central pixel \\
\hline
\end{tabular}
\end{table}\label{line_sheet}

  \section{Method}\label{sect:method}
As described in \cite{Wang+2017}, {\it fiDrizzle-DC} makes a noteworthy balance between image fidelity, noise control and convergence speed. As a reasonable generalization, in circumstances where all the pixel values in the input images are positive, we could perform a multiplicative update to refresh each iterative result. The workflow of this generalization, i.e. the {\it fiDrizzle-MU} algorithm is described below:

\begin{enumerate}

\item Given $N$ dithered exposures of the same imaging field, $\{I^1, I^2... I^N\}$, as input, {\it Drizzle} them onto a finer grid to produce a high resolution image $F_0$.The subscript denotes the order of the iteration. 

\vspace{0.2cm}

\item Map the result of the $0-$th iteration, $F_0$, back to the same $N$ frames of the input images, to simulate those dithered exposures, producing a set of approximations to the observation, $G^k_0$.

\vspace{0.2cm}

\item Divide the original images by their corresponding approximations to yield a set of ratio images $R^k_1 = \frac{I^k}{G^k_0}$, sharing the same frames as the dithered ones. As the term literally implies, $R$ denotes a ratio.

\vspace{0.2cm}

\item Return to the first step and now take the new image set $\{R^1_0, R^2_0...R^N_0\}$ as input to Drizzle into a new image $R_1$ on the finer grid. And a pixel-wise normalization is conducted to $R_1$ by the effective overlapping layers $L_E$.  

\vspace{0.2cm}

\item Continue as before and keep in mind that for $i\geq1$, the direct product of $Drizzle$ is $R_i$ which is a multiplicative correction to update the $(i-1)-$th reconstruction, while the $i-$th one, $F_i$, is obtained by multiplying $R_i$ by $F_{i-1}$. 

\end{enumerate}

A reminder is warranted that the up-sampling operation, i.e. $Drizzle$, employed in the above steps operates with a $pixfrac$ value of 1, which in practice, corresponds to the {\it shift-and-add} technique. The down-sampling process from fine-grid to coarse-grid pixels is implemented through an area-weighted flux conservation scheme, where the value of each coarse pixel is computed as the integration of overlapping contributions from all intersected fine-grid pixels. The convergence rate can be algorithmically modulated by elevating the multiplicative update to the $\gamma$-powered step size parameter, where $\gamma$ defines the exponential scaling factor for iterative refinement. In this work, we intentionally set $\gamma=1$ as the baseline value, a conservative yet theoretically justified choice to ensure algorithm stability across all test conditions. For the sake of clarity, we denote $\mathfrak{S}^k_u$ to signify the up-sampling operation from the $k$-th coarse frame to the fine one and $\mathfrak{S}^k_d$ to signify the corresponding backward down-sampling operation (see the variables in Table \ref{line_sheet}). The complete algorithm can subsequently be expressed mathematically as follows:

\begin{equation}\label{fiD-MU}
F_{i+1} = F_{i} \times \Bigg{(} \frac{1}{L_E} \sum_{k=1}^{N} \mathfrak{S}^k_u \left\{ \frac{I^k}{\mathfrak{S}^k_d\{ F_i \} } \right\} \Bigg{)}^{\gamma}.
\end{equation}

\vspace{0.2cm}


By applying multiplicative updates, {\it fiDrizzle-MU} converges to the optimal value faster than {\it fiDrizzle-DC}. In general, the number of steps required for convergence depends on the number of dithered frames and the {\tt PSR} parameter. Intuitively, employing a larger number of dithered frames in conjunction with smaller {\tt PSR} parameters necessitates more iterative steps to achieve optimal reconstruction. Additionally, stronger noise backgrounds extends the iterations needed for convergence. And with the intention of attenuating high-frequency artifacts induced by the iterative processing such as the ringing effects, we simply apply a non-strict positivity constraint to the reconstructions, which serves as a standard regularization term frequently employed in image combination and signal processing applications:

\begin{equation}\label{constraints}
F_{i+1}=
\begin{cases}
 F_{i+1},&F_{i+1} \ge 0\\
 F{i},&F_{i+1} < 0
\end{cases}
\end{equation}

  \section{Analysis}\label{sect:analy}
\subsection{Dither, sampling and pixel correlation}\label{dsc}
We begin our exploration by examining the procedure for producing a high-resolution image with Drizzle from the true brightness distribution of a light source which is denoted by {\it $\ L_t$}. This process involves a series of operations such as PSF convolution, dithering, CCD sampling and resampling (all conducted without considering noise in this analysis). 

Before reaching the detector, which is typically a CCD array, {\it $\ L_t$} is convolved by the telescope's optics, transformed into a light distribution blurred by the telescope's PSF,  {\it $ \ L_b = L_t \otimes K_o$} where {\it $K_o$} represents the PSF and {\it $L_b$} denotes the PSF-blurred light distribution.  Multiple exposures with sub-pixel dithering are widely used to attain superior image quality in astronomical observations. This strategy increases the effective integration time, improving the signal-to-noise ratio, i.e. {\it SNR}, and achieves high-frequency sampling of the target sources to obtain additional spatial information, while minimizes the risk of saturation in a single long exposure. As the dithered images are commonly Drizzled onto a fine pixel grid, the objective of the reconstruction is to recover the underlying high-definition image sampled on that fine grid --- {\it $L_f = L_b \otimes P_f$}, which is obtained by sampling {\it $L_b$} on the target fine pixel grid {\it $P_f$}. Every single exposure captures an image by sampling {\it $L_b$} with the detector, where {\it $L_b$} is convolved with the electronic pixels {\it $P_e$} of the camera in concordance with that dithered frame, resulting in a pixelated image formulated by {\it $I^k = W^k \times L_b \otimes P_e^k$} with the weight map {\it $W^k$} taken into account. The superscript notation is used to distinguish the {\it k-}th frame in the dithered sequence. Following the above analysis, {\it $L_a$} serves as an approximation to {\it $L_f$} in the form of:

\begin{equation}\label{sampling}
\begin{aligned} L_{a} & =\sum_{k=1}^{N} W^k \times L_{t} \otimes K_{0} \otimes P_{e}^{k} \otimes P_{f} \\ & =L_{t} \otimes K_{0} \otimes \sum_{k=1}^{N} \Bigg{(}W^k \times P_{e}^{k}\Bigg{)} \otimes P_{f} \end{aligned}
\end{equation}

As shown in equation \ref{sampling}, when a high-definition image (no matter it is an idealized image with infinitely high resolution or its sampled representation on a fine-pixel grid) is convolved with $P_{e}^{k}$, it will be forwardly sampled onto the $k-$th coarse-pixel dithered frame. In contrast, the adjoint operator $P_{e}^{k*}$ performs the reverse operation. The fact that convolution obeys the associative and distributive properties enables the summation sign to be moved ahead of the term inside the parentheses in the second equation. Consequently, we are able to formulate a convolution kernel that encapsulates the combined impact of dithering, resampling, and coaddition operations, which can literally be referred to as the dithering kernel {\it $K_d \ $} :

\begin{equation}\label{dithering_kernel}
K_d = \sum_{k=1}^{N} \Bigg{(}W^k \times P_{e}^{k} \Bigg{)}
\end{equation}

\vspace{0.2cm}

The two equations stated above indicate that, from a mathematical standpoint, the high-resolution image obtained through Drizzling the dithered multi-exposure images is the one calculated by convolving  the intrinsic PSF-blurred image {\it $L_b$} with {\it $K_d$} (which must be normalized) then sampling in the high-resolution fine grid, which is analogous to the blurring effect caused by a PSF. This provides a clear understanding of the spatial correlation between a pixel and its neighboring pixels, as seen in the Drizzled image. Or from an alternative perspective, the issue of removing inter-pixel correlations in a Drizzled image is essentially equivalent to a deconvolution problem. An interesting property of $K_d$, as seen from its expression, is that it can be interpreted as the sum of $N$ sub-kernels, with the {\it k-}th sub-kernel, $\mathscr{K}_d^k$, is simply $P_e^k$. And the adjoint operator of $\mathscr{K}_d^k$, namely $\mathscr{K}_d^{k*}$, corresponds to $P_e^{k*}$. Specifically, each dithered frame contributes its own sub dithering kernel, and $K_d$ represents the weighted average of these sub-kernels, with weights given by $W^k$, which highlights the role of individual frame contributions in shaping the final convolution kernel.

\begin{figure}
\centering
\includegraphics[width=5.7in]{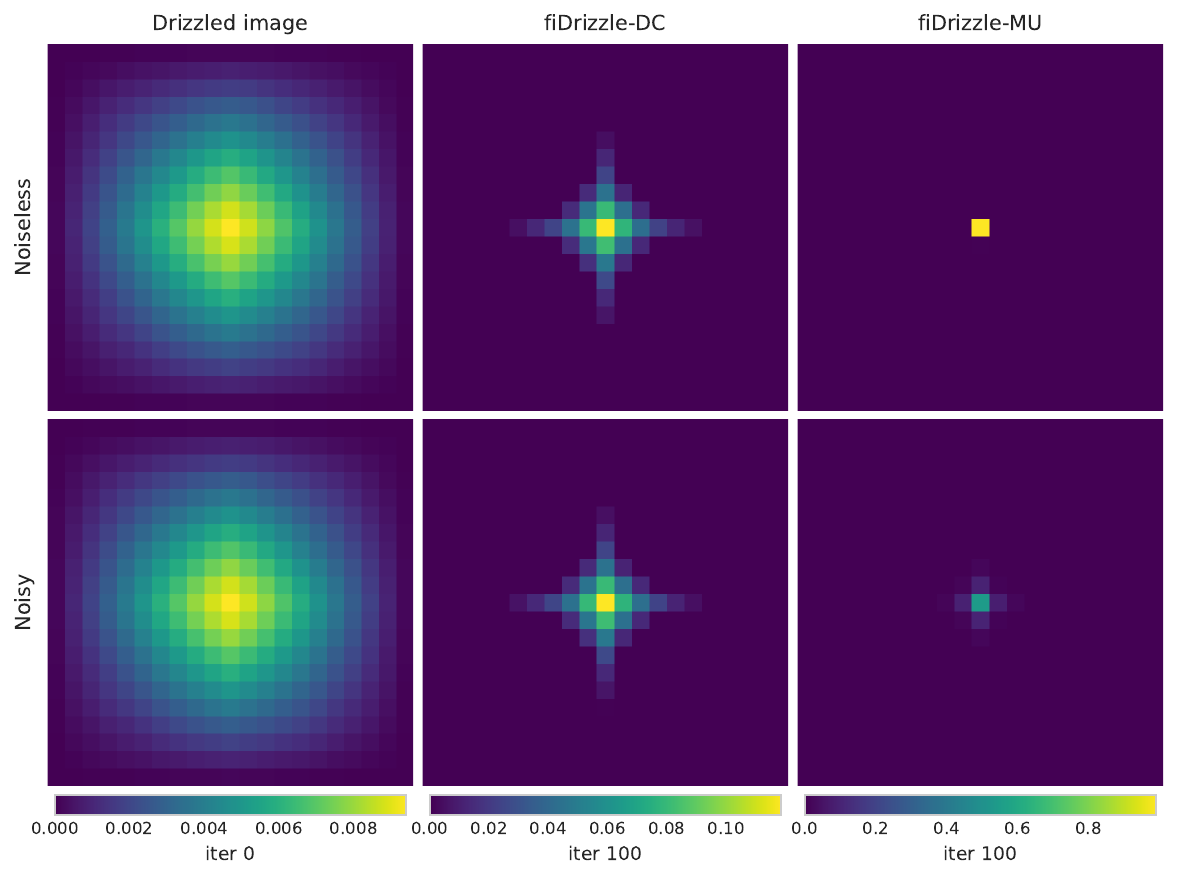} 
\caption{Results of 1,600 dithered simulated observations of a point source located on a fine grid with $0.005^{\prime \prime}$ per pixel. Each simulated observation is sampled at the CSST-MCI pixel scale of $0.05^{\prime \prime}$, with Poisson noise corresponding to a mean background level of 500 counts per pixel applied prior to normalization. The total flux of the point source is likewise set to 500 counts before normalization. The left column shows the combined image obtained by Drizzling all 1,600 mock observational images back onto the $0.005^{\prime \prime}$ grid. The middle column presents the reconstruction result obtained with {\it fiDrizzle-DC} after 100 iterations, while the right column shows the corresponding result from {\it fiDrizzle-MU}. The top row corresponds to noise-free simulations, while the bottom row include Poisson noise.} \label{fig_iter100}
\end{figure}

The explicit representation of $K_d$ is typically governed by the Drizzle parameters, such as the {\tt PSR} parameter and the dither pattern used during the observations. For an ideal dither pattern consisting of infinite number of uniformly distributed random sub-pixel shifts, {\it $K_d$} can be computed in an almost analytical form. Furthermore, due to the pixel-scale translational symmetry of the dither pattern in the internal region of the image, different pixels in this region will share the same dithering kernel. Figure \ref{fig_kernel} illustrates the dithering kernel for a {\tt PSR} parameter of 0.1, with the top-left panel depicting the ideal dithering kernel as described earlier, while the remaining panels show the results for a finite number of dithered frames visualized in terms of the pixel-wise relative residuals with respect to the theoretical kernel. Figure \ref{fig_iter100} (left column) presents dithering kernels obtained from 1600 dithered frames. The top and bottom panels illustrate the noise-free and Poisson noise cases, respectively, both of which exhibit a visual resemblance to the theoretical kernel. It is clear that, with a sufficiently large number of dithered frames, the pattern obtained from Drizzling a latent point closely matches the theoretical dithering kernel, with the differences diminishing as the number of dithered frames increases. A reliable image reconstruction method should efficiently deconvolve the dithering kernel, focusing the spread energy back to the central point.

In reality, the flux redistribution after Drizzle will be affected by the random noise imported during the sampling process, deviating from the noiseless dithering kernel. Nevertheless, the correlated noise between different pixels in the Drizzled image will retain a structure similar to the dithering kernel. While this is an ill-posed problem, a well-designed image reconstruction method can still recover the flux in the target pixel grid sufficiently.

\subsection{Flux re-concentration capability}
A bright spot with a total flux of 500 counts was dithered 1,600 times on a fine pixel grid with a resolution of $0.005^{\prime \prime}$, and then down-sampled to the coarser CSST-MCI resolution of $0.05^{\prime \prime}$ to generate a group of mock images. These images were subsequently Drizzled to produce a combined image with a pixel scale of $0.005^{\prime \prime}$ and normalized by the aggregated flux of this point source. This procedure partially follows the same process as that described in subsection \ref{dsc}. Following this, we used the normalized Drizzled image as an initial input for iterative processing with {\it fiDrizzle}. In addition, we generated mock images incorporating Poisson noise. Before normalization, each mock image was assigned a Poisson background with a mean of 500 counts per pixel, matching the total flux of the point source. This configuration represents a highly noisy condition, providing an opportunity to evaluate the effectiveness of different reconstruction methods in accurately recovering point source fluxes in the presence of severe noise. 

Figure \ref{fig_iter100} provides an illustrative example of the outcome after 100 iterations of {\it fiDrizzle} applied to the Drizzled image. In this figure, the middle column demonstrates the reconstructions obtained by {\it fiDrizzle-DC}, whereas the right column shows the comparable results of {\it fiDrizzle-MU}. As shown, the initially diffused flux pattern in the Drizzled image is substantially gathered back towards the central region, corresponding to the original point source. And it is clearly distinguishable that {\it fiDrizzle-MU} outperforms {\it fiDrizzle-DC} regardless of the presence of noise, leading to a more substantial accumulation of flux towards the central region, at least by the 100-th iteration. When Poisson noise is present, the rate at which the flux is concentrated is somewhat slower relative to the noise-free case when {\it fiDrizzle-MU} is employed. This process of centripetal flux concentration can be interpreted as the result of deconvolving the corresponding dithering kernels shown on the left side of the figure.

\begin{figure}
\centering
\includegraphics[width=5.5in]{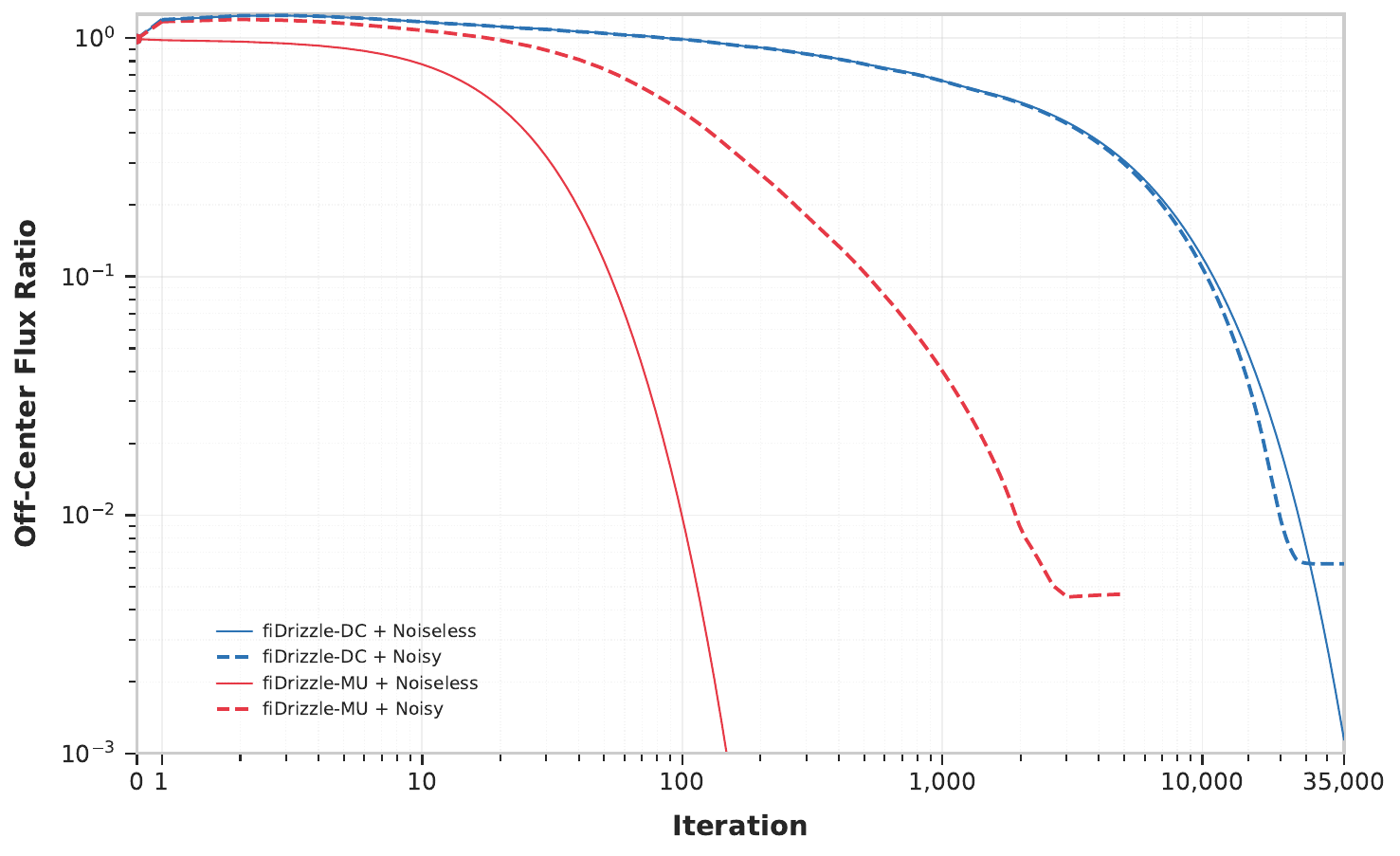} 
\caption{Evolution of the OCFR, defined as the normalized sum of pixel values excluding the central pixel, versus iteration number for {\it fiDrizzle-DC} and {\it fiDrizzle-MU} algorithms. Both axes are plotted on a logarithmic scale. The horizontal axis shows iteration steps, starting from iteration 1. The point at iteration “0”, representing the initial Drizzled image, is explicitly added for reference, as zero is not defined in log space. Solid lines correspond to noiseless cases, dashed lines indicate Poisson noise cases. Red curves represent {\it fiDrizzle-MU}, blue curves {\it fiDrizzle-DC}. In the noiseless case,  achieves zero OCFR by iteration 850 (truncated after $\sim$150 steps for clarity). For the noisy case, {\it fiDrizzle-MU} results are shown up to 5000 iterations, whereas {\it fiDrizzle-DC}, exhibiting slower convergence, is iterated up to 35,000 steps in both noise conditions.} \label{fig_ocfr}
\end{figure}


In order to provide a quantitative characterization of the ability of different reconstruction methods to recover the aliased pixel intensities introduced by dithering and \text{Drizzle} --- specifically, their capacity to decorrelate neighboring pixels --- we calculate two primary metrics: (1) the OCFR, defined as the sum of pixel values excluding the central pixel normalized by the total flux, providing a reliable quantification of the correlation strength between a given pixel and its neighbors in the combined image --- a lower OCFR indicates better flux concentration and stronger pixel decorrelation, and (2) the PSNR, measuring the overall reconstruction fidelity --- higher PSNR values correspond to reconstructions that more closely match the ground truth, indicating better preservation of image details and lower reconstruction error. 

Figure \ref{fig_ocfr} provides a detailed comparison of the OCFR for the two reconstruction methods, {\it fiDrizzle-DC} and {\it fiDrizzle-MU}, under both noiseless and Poisson noise conditions. In this figure, both axes are plotted on a logarithmic scale to capture the convergence behavior over several orders of magnitude. As shown, {\it fiDrizzle-MU} demonstrates a significantly faster convergence rate compared to {\it fiDrizzle-DC}. In the noiseless case, the OCFR of {\it fiDrizzle-MU} rapidly decreases and reaches zero by iteration 850. Beyond this point, there is no further decrease in OCFR, as the flux has been fully reconcentrated at the central pixel. For clarity, the noiseless {\it fiDrizzle-MU} curve is truncated after approximately 150 iterations in the plot, since the subsequent values decline sharply. Under Poisson noise conditions, {\it fiDrizzle-MU} maintains superior performance relative to {\it fiDrizzle-DC}. In this case, {\it fiDrizzle-MU} is iterated up to 5000 steps to demonstrate its continued convergence, albeit at a reduced rate compared to the noiseless scenario. In contrast, {\it fiDrizzle-DC} exhibits much slower convergence behavior in both noiseless and noisy cases, while it appears to be less affected by noise, with the OCFR values remaining consistent until $\sim$30000 iterations regardless of the presence of noise. However, even after 35,000 iterations — the maximum number of iterations tested in this analysis — {\it fiDrizzle-DC} fails to achieve an OCFR comparable to that of {\it fiDrizzle-MU} in noisy conditions. This suggests that {\it fiDrizzle-MU}, by employing the multiplicative update, is inherently more robust and efficient in addressing the flux redistribution introduced by the dithering and Drizzle processes. These results highlight the superior ability of {\it fiDrizzle-MU} to rapidly decorrelate pixels from their neighbors, thereby achieving a more accurate reconstruction in both ideal and realistic noisy scenarios.


\begin{figure}
\centering
\includegraphics[width=5.7in]{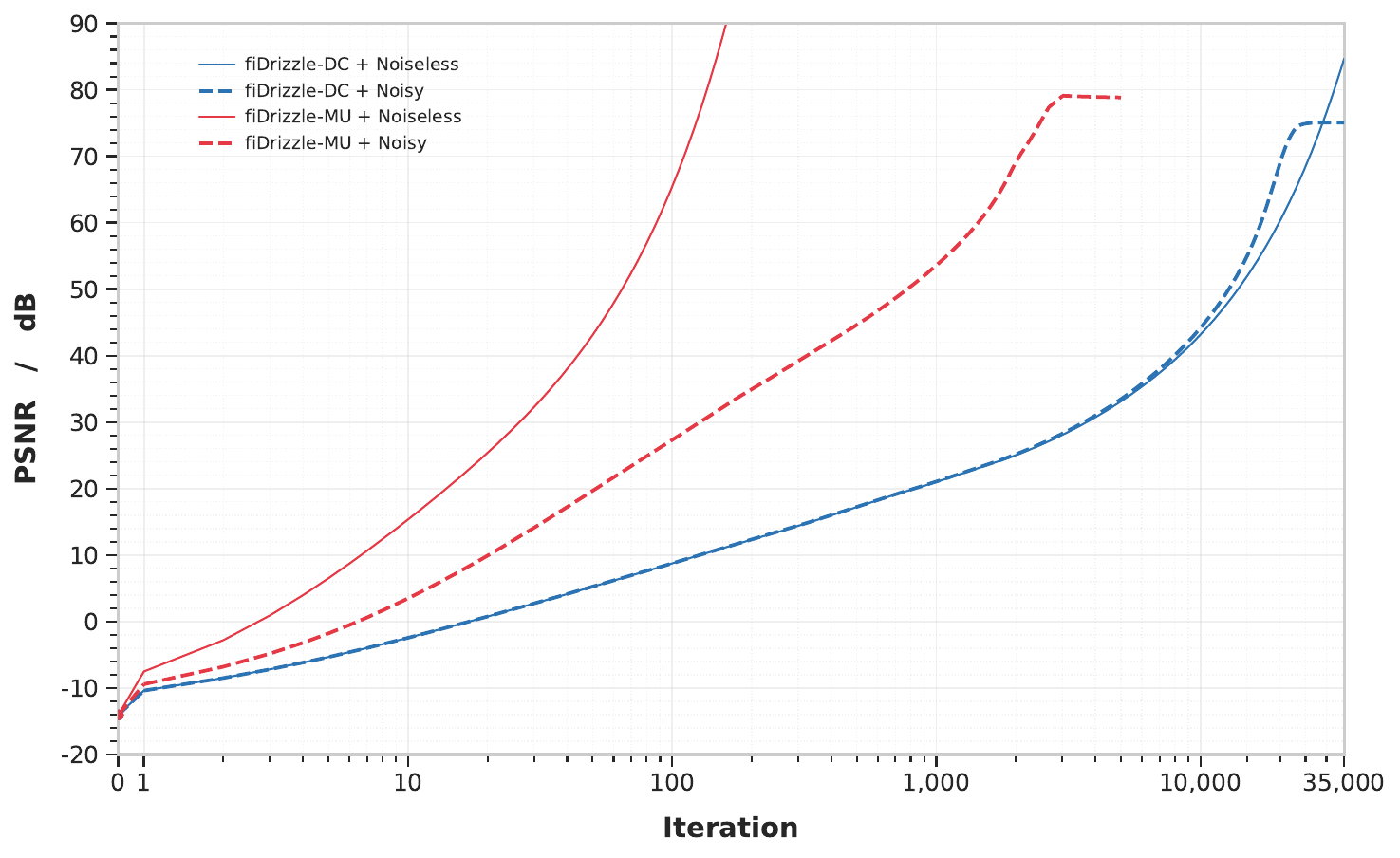} 
\caption{Evolution of the PSNR versus iteration number for {\it fiDrizzle-DC} and {\it fiDrizzle-MU}. Plotting conventions (logarithmic axes, line styles, color coding, and iteration steps in each condition) are identical to those in Figure~\ref{fig_ocfr}. The PSNR of the $0-$th iteration, i.e. the \textit{Drizzle} result, is -14.035 dB. Similar trends are observed: {\it fiDrizzle-MU} exhibits significantly faster convergence in the noiseless case (curve truncated after $\sim$150 steps), and maintains superior reconstruction fidelity under Poisson noise conditions compared to {\it fiDrizzle-DC}.} \label{fig_psnr}
\end{figure}

In addition to OCFR, we further assess the reconstruction quality using the PSNR between the reconstructed images and the ground truth. As shown in Figure~\ref{fig_psnr}, the PSNR exhibits trends similar to those of the OCFR metric. Specifically, {\it fiDrizzle-MU} demonstrates a much faster improvement in reconstruction fidelity compared to {\it fiDrizzle-DC}, under both noiseless and noisy conditions. In the noiseless case, {\it fiDrizzle-MU} achieves a PSNR exceeding 80 dB within 130 iterations, indicating a high-fidelity reconstruction. Even in the presence of Poisson noise, {\it fiDrizzle-MU} reaches a PSNR of 76 dB after about 2,550 iterations. In stark contrast, {\it fiDrizzle-DC} requires nearly 30,000 iterations to approach similar PSNR levels in the noiseless scenario. Furthermore, under noisy conditions, {\it fiDrizzle-DC} fails to achieve comparable PSNR values, regardless of the number of iterations performed. These results clearly demonstrate the superior convergence rate and reconstruction accuracy of {\it fiDrizzle-MU}, particularly in challenging noise-dominated environments.


\begin{figure}
\centering
\includegraphics[width=5.7in]{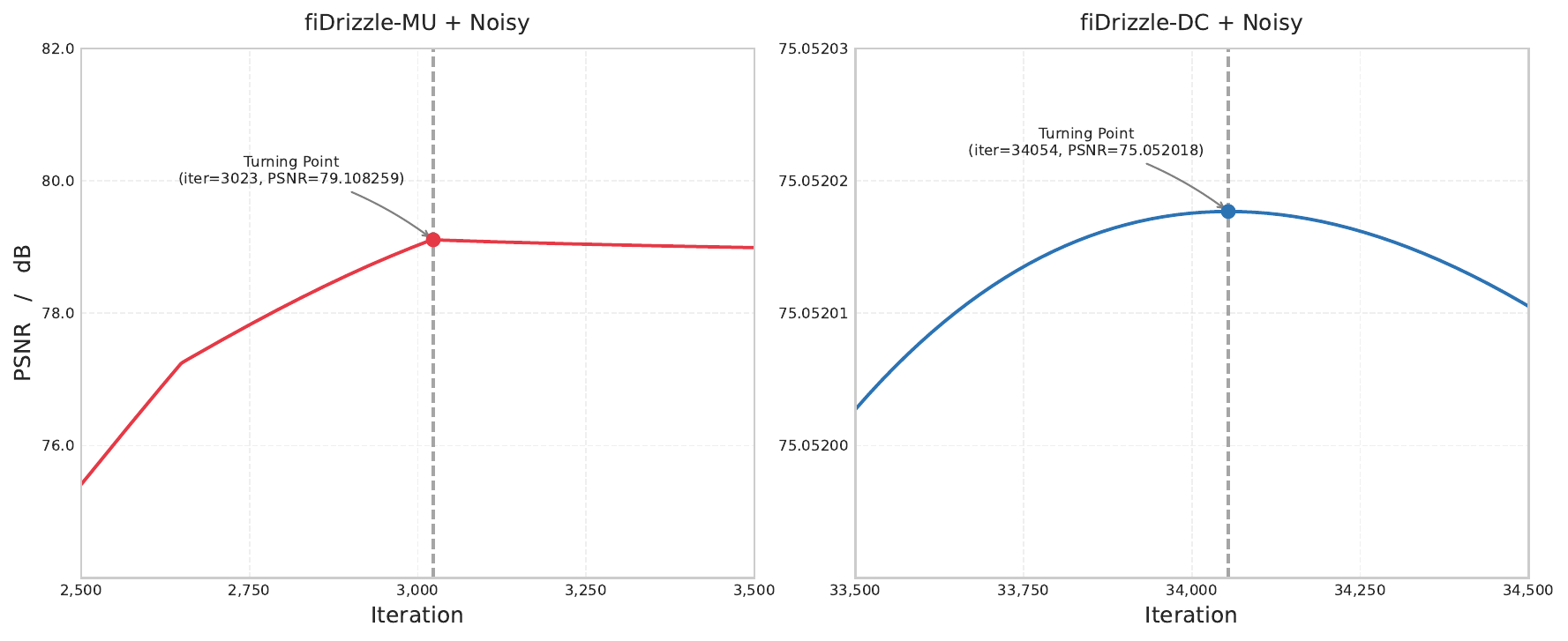}
\caption{Evolution of the PSNR with iteration number, demonstrating the turning point behavior in iterative reconstructions under Poisson noise. The {\it fiDrizzle-MU} algorithm achieves its maximum PSNR of 79.11 dB at iteration 3023, beyond which the PSNR decreases due to noise amplification. Similarly, {\it fiDrizzle-DC} reaches its maximum PSNR of 75.05 dB at iteration 34054. This phenomenon is consistent with the typical characteristics of iterative deconvolution methods, where the iteration number serves as an implicit regularization parameter controlling the balance between reconstruction fidelity and noise suppression.} \label{fig_psnr_turning_points}
\end{figure}

Since we regard the process of mitigating the degradation caused by dithering and Drizzle as a deconvolution problem, the reconstruction exhibits characteristics similar to other iterative deconvolution processes. In particular, the iteration number serves as an implicit regularization parameter, governing the trade-off between enhancing resolution and amplifying noise. Figure~\ref{fig_psnr_turning_points} illustrates the evolution of the PSNR as a function of iteration number for both {\it fiDrizzle-MU} and {\it fiDrizzle-DC} around their respective optimal results under Poisson noise conditions. As the number of iterations increases, the PSNR initially rises, reflecting improved reconstructions. However, after reaching a maximum, the PSNR begins to decline, indicating the onset of overfitting and noise amplification. For {\it fiDrizzle-MU}, the turning point occurs at iteration 3023, where the PSNR attains its peak value of 79.108 dB. Beyond this point, further iterations result in a gradual degradation of reconstruction quality, as reflected by the decreasing PSNR. A similar trend is observed for {\it fiDrizzle-DC}, although its turning point is delayed, occurring at iteration 34054 with a maximum PSNR of 75.052 dB. After this iteration, continued iterations lead to a slow but consistent decline in PSNR. These results highlight the importance of controlling the number of iterations in iterative reconstruction algorithms, as excessive iterations can compromise reconstruction fidelity by amplifying noise. This phenomenon is consistent with the conclusions of \citealt{BB2001} and other studies on iterative deconvolution methods. Before processing observational data, we typically perform simulations on the data to determine the optimal iteration step at which to stop for achieving the best solution.

These results confirm that the multiplicative update strategy adopted by fiDrizzle-MU not only accelerates convergence but also improves the overall reconstruction accuracy, particularly in the presence of observational noise.

\subsection{The positivity constraints}
As mentioned in Equation \ref{constraints}, positivity constraints are incorporated into our iterative scheme to enforce the physical requirement that flux values in astronomical images must be non-negative. In the context of image reconstruction and deconvolution, positivity serves as an important prior, preventing the algorithm from introducing unphysical negative flux values that commonly arise due to noise amplification or overfitting during iterations \citep{BB2001}. Positivity constraints, along with other constraints such as band limiting and wavelet transform, act as a form of regularization, suppressing high-frequency oscillations and stabilizing the convergence process \cite{Starck+1994}. Without such constraints, iterative algorithms often produce ringing artifacts and spurious negative sidelobes, especially when recovering high-frequency features is essential \citep{Starck+2002}. These artifacts can significantly degrade the photometric and morphological accuracy of the recovered sources.

\begin{figure}
\centering
\includegraphics[width=5.5in]{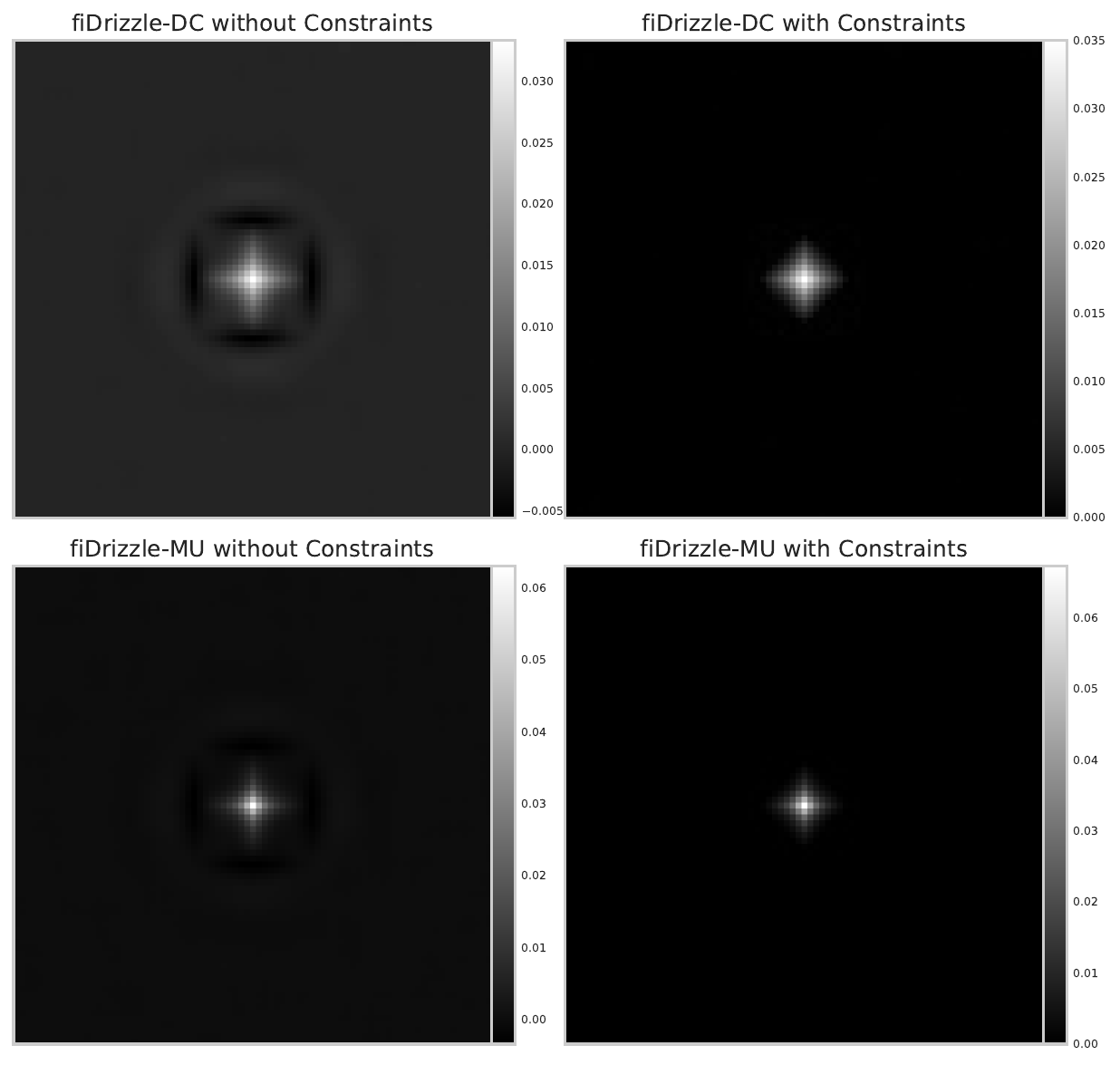}
\caption{Comparison of reconstruction results after 10 iterations of {\it fiDrizzle-DC} and {\it fiDrizzle-MU}, with and without positivity constraints. The left panels show the reconstructions without positivity enforcement, while the right panels display the results with positivity constraints applied. Without the constraints, both algorithms exhibit noticeable ringing artifacts and spurious negative flux values. In contrast, applying positivity constraints significantly suppresses these artifacts, leading to cleaner reconstructions and better flux localization.}
\label{fig_constr}
\end{figure}

To evaluate the effect of positivity constraints in our framework, we compare the results of {\it fiDrizzle-DC} and {\it fiDrizzle-MU} after 10 iterations, both with and without positivity enforcement. The results are presented in Figure \ref{fig_constr}. In the absence of positivity constraints (left panels), both algorithms exhibit pronounced ringing effects around bright sources, as well as noticeable negative flux regions. These artifacts distort the reconstructed image and reduce the dynamic range. While with the positivity constraints applied,these high-frequency artifacts are effectively suppressed. 

Recalling the description of the {\it fiDrizzle-MU} algorithm in Section \ref{sect:method} and the analysis of the dithering kernel---particularly the sub-kernels---in Subsection \ref{dsc}, {\it fiDrizzle-MU} can, in fact, be formally expressed as:

\begin{equation}\label{fiDrizzle-MU2R&L}
F_{i+1}=F_{i} \times\left(\frac{1}{L_{E}}\left\{\sum_{k=1}^{N} \frac{I^{k}}{W^{k} \times \mathscr{K}_d^k \otimes F_{i}} \otimes \mathscr{K}_d^{k *}\right\}\right)
\end{equation}

This formulation indicates that {\it fiDrizzle-MU} can be interpreted as performing a Richardson-Lucy (RL)-like deconvolution  (\cite{Richardson+1972}, \cite{Lucy+1974}) for each individual dithered frame, using its corresponding sub-dithering kernel $\mathscr{K}_d^k$. Specifically, the algorithm iteratively deconvolves the blurring effects introduced by the sub dithering kernel in each dithered frame. At each iteration, the deconvolved contributions from all $N$ dithered frames are aggregated by weighted summation. The reconstructions thus incorporate not only the additional fine-scale details sampled differently across the multiple observations, but also benefit from the averaging of noise components present in each individual observation. This averaging effect leads to improved noise suppression and results in a smoother and more stable reconstruction at each iteration step. This RL-like scheme, when combined with positivity constraints and appropriate stopping criteria, provides a flexible yet robust framework for high-fidelity image reconstruction from dithered observations. This formulation corresponds to Equation (20) in \cite{Starck+2002}, and it is particularly well-suited for astronomical imaging, where Poisson noise is the predominant noise component.

Similarly, {\it fiDrizzle-DC} can be regarded as a Landweber-type iterative algorithm \citep{Landweber+1951}. When a positivity constraint is introduced, the method corresponds to Equation (15) in \cite{Starck+2002}. Consistent with the case of PSF deconvolution, the convergence speed and reconstruction fidelity of the Landweber-type algorithms are generally inferior to those of the RL-type ones.

\subsection{The spatially resolving power}\label{subsect:JWST}

\begin{figure}
\centering
\includegraphics[width=\textwidth]{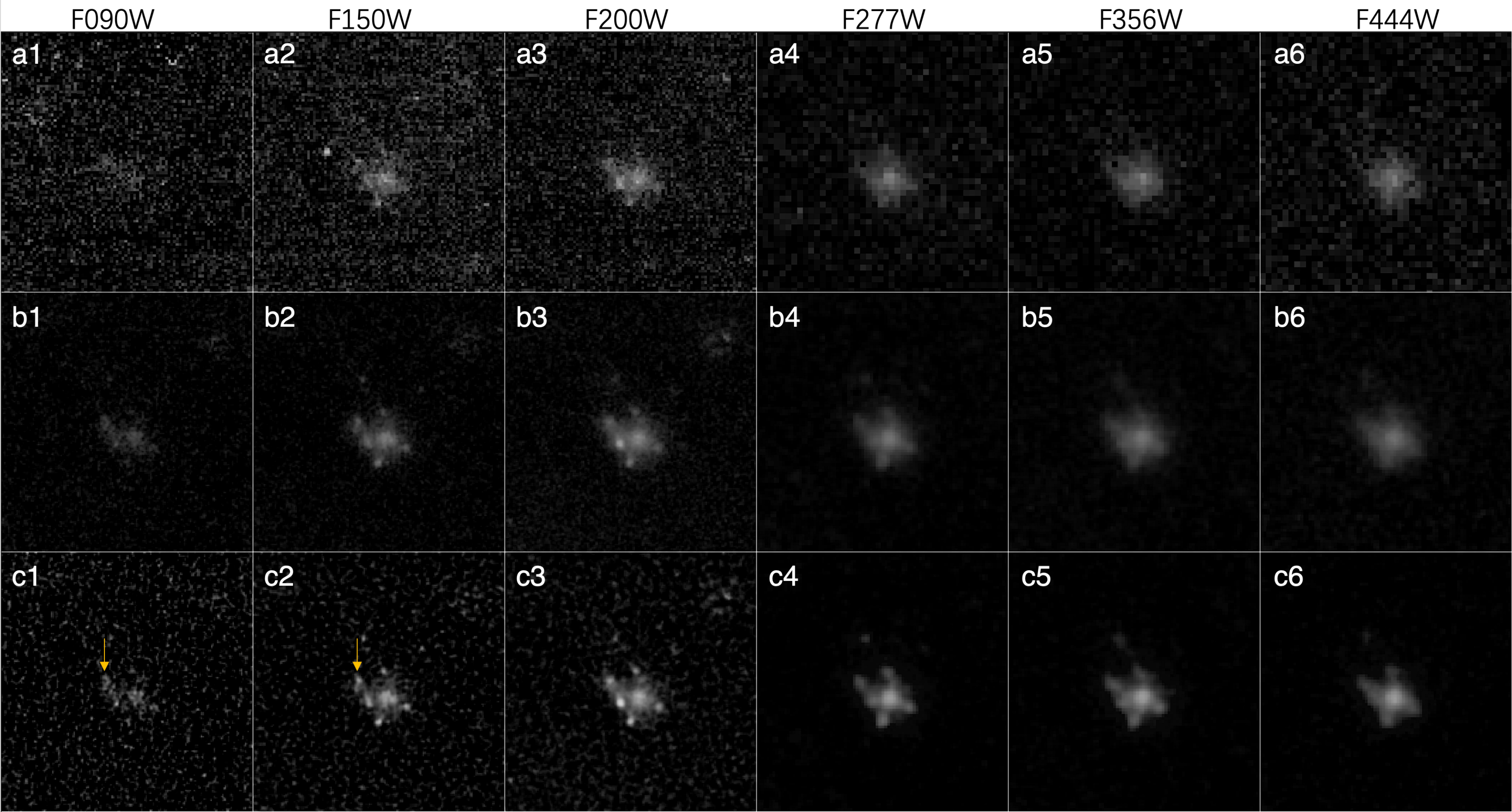}
\caption{ Anti-aliasing a gravitationally lensed quasar candidate captured by JWST-NIRCam in six bands: three short-wavelength bands F090W, F150W, F200W and three long-wavelength bands F270W, F356W, F444W. The top row (a1-a6) presents the "*cal.fits" observational images calibrated in stage 2 of the JWST pipeline. The middle row (b1-b6) features panels generated by the \textit{Drizzle} algorithm in stage 3 (with 9 exposures coadded for each panel), while the bottom row (c1-c6) displays results processed with the \textit{fiDrizzle-MU} method, where c1-c3 did not apply the positive constraint from Eq. \ref{constraints}, while c4-c6 did. The candidate locates at $R.A.=110^\circ.761440, DEC=-73^\circ.452486$, nearby the strong lensing system SMACS J0723.3-7327.}
\label{fig_JWST1}
\end{figure}

Figure \ref{fig_JWST1} illustrates images of a candidate gravitationally lensed quasar system captured by JWST-NIRCam across six distinct bands: three short-wavelength bands F090W, F150W, F200W and three long-wavelength bands F270W, F356W, F444W. 

In the top row (a1-a6), the "*cal.fits" observational images are presented, having been calibrated through stage 2 of the JWST pipeline. The pixel scale of the observational images differs between the short-wavelength and long-wavelength bands for NIRCam on JWST. Specifically, in the short-wavelength bands, the pixel size is 0.031\arcsec/pixel, whereas in the long-wavelength bands, it is 0.063\arcsec/pixel. This variation in pixel scale is designed to optimize the performance and resolution of the instrument across different wavelengths. 

The middle row (b1-b6) displays the panels produced by the \textit{Drizzle} algorithm (with 9 exposures coadded for each panel) during the stacking process in stage 3 of the JWST pipeline. The bottom row (c1-c6) showcases results processed using the \textit{fiDrizzle-MU} method, where images c1-c3 did not apply the positive constraint detailed in Eq. \ref{constraints}, whereas images c4-c6 adhered to this constraint. The positive constraint has visibly reduced the ringing effects around bright sources, thereby achieving better control over the background noise. This improvement helps in distinguishing true signals from noise, offering clearer images and more accurate data interpretation, especially in the context of complex systems like gravitationally lensed quasars. By mitigating the ringing effect, the analysis and identification of such phenomena are significantly enhanced, allowing for more precise scientific conclusions to be drawn from the observations. The panles in the middle and bottom rows have been upsampled by a factor of {\tt PSR=0.5} compared to the images in the top row. This means that the pixel scales for the middle and bottom rows are halved. Specifically, for the short-wavelength bands, the pixel size is 0.0155\arcsec/pixel, and for the long-wavelength bands, it is 0.0315\arcsec/pixel.

The point-like source indicated by the yellow arrow demonstrates significant flux across all six bands, while the other four point-like sources surrounding the lens show notable flux only in bands c2-c6. This suggests that these four sources may originate from the same quasar (QSO) located behind the lensing galaxy, implying that the source highlighted by the yellow arrow is not an image produced by the lensing effect.

Notably, prior to this data processing, we conducted 30 simulations of the source and noise conditions in this field of view to determine the number of iterations that would yield the highest PSNR. The optimal range was identified to be between 55 and 75, and we ultimately selected 65 iterations.

\section{Computational consumption}\label{sect:comput}

We conducted three sets of simulations—Mock-I, Mock-II, and Mock-III—to evaluate the computational cost of four image stacking methods: {\it Drizzle}, {\it iDrizzle}, {\it fiDrizzle-DC}, and {\it fiDrizzle-MU}. The specifics of each simulation are as follows:

\begin{enumerate}

\item Mock-I: Stacked 10 exposure images sized at $512\times512$ pixels with a sampling parameter {\tt PSR} = 0.5.
\item Mock-II: Stacked 80 exposure images sized at $256\times256$ pixels with a sampling parameter {\tt PSR} = 0.25.
\item Mock-III: Stacked 160 exposure images sized at $128\times128$ pixels with a sampling parameter {\tt PSR} = 0.1.
\end{enumerate}
The time consumption for these processes is listed in Table \ref{Tcomput}, where the numbers in parentheses indicate the number of iterations required to achieve the same PSNR, set here at PSNR = 20. It's noted that {\it iDrizzle}, {\it fiDrizzle-DC}, and {\it fiDrizzle-MU} all employed a positive value constraint during the image stacking process.

It was observed that {\it fiDrizzle-MU} and {\it fiDrizzle-DC} consumed approximately the same amount of time per iteration, roughly equivalent to twice the time taken by {\it Drizzle}. However, due to its higher convergence rate, {\it fiDrizzle-MU} achieved the same results in about one-fifth of the time compared to {\it fiDrizzle-DC}. The {\it iDrizzle} algorithm, which frequently calls Fast Fourier Transform (FFT) for filtering during iterations, took approximately six times longer than {\it fiDrizzle-DC}.

This comparative analysis highlights the efficiency and effectiveness of different image stacking techniques under various conditions, with {\it fiDrizzle-MU} showing significant advantages in terms of computational efficiency and convergence speed when reaching a specified PSNR.

 \begin{table}
    \caption{Computation consumed by various approaches to the same PSNR.}\label{computationtime}
        \centering
        \begin{tabular}{crrrr}
         \hline
        {\bf{Simu No.}} & {\it Drizzle} & {\it iDrizzle}    & {\it fiDrizzle-DC}  & {\it fiDrizzle-MU}   \\
        \hline
        Mock-I ($512\times512\times10,{\tt PSR}=0.50$) & 1.2s & 2407.1s(164) & 267.9s(123)  & 66.8s(27)   \\
        Mock-II ($256\times256\times80,{\tt PSR}=0.25$) & 4.9s & 9490.8s(159) & 1068.8s(119) & 242.1s(24)   \\
        Mock-III ($128\times128\times160,{\tt PSR}=0.10$) & 30.3s & 56265.7s(155) & 6190.9s(114) & 1410.6s(22)  \\
        \hline
        \end{tabular}
        \label{Tcomput}
    \end{table}

\section{Discussion and Conclusion}\label{sect:discu&conclu}

In this work, we have proposed and evaluated {\it fiDrizzle-MU}, an enhanced iterative algorithm for the reconstruction of high-resolution images from dithered astronomical observations. By replacing the difference-based correction term in {\it fiDrizzle-DC} with a multiplicative update, and incorporating positivity constraints, {\it fiDrizzle-MU} demonstrates significant improvements in both convergence speed and reconstruction fidelity. Through comprehensive numerical experiments, we have shown that the algorithm effectively mitigates aliasing and deblends flux dispersed by the Drizzling process.

Our analysis in Section \ref{sect:analy} confirms that {\it fiDrizzle-MU} outperforms its predecessor, {\it fiDrizzle-DC}, across several key metrics. The proposed method accelerates the concentration of dispersed flux back to its source locations, as evidenced by the rapid decline in the OCFR. Additionally, the PSNR analysis highlights its superior ability to recover intrinsic image structures, even in the presence of significant Poisson noise. Importantly, the algorithm maintains robustness against noise amplification through the integration of positivity constraints, which suppress spurious oscillations and ringing artifacts commonly associated with iterative deconvolution methods. In another respect, the quantitative analysis presented in Section \ref{sect:comput} clearly demonstrates the substantial advantage of {\it fiDrizzle-MU} in reducing computational consumption --- this renders it especially suitable for combining CSST-MCI Extremely Deep Field data, which comprises 1,600 dithered exposures per pointing.

Moreover, we have demonstrated in Subsection \ref{subsect:JWST} that {\it fiDrizzle-MU} significantly enhances spatial resolution, resolving structures with minimal separations that were otherwise blended in conventional reconstructions. This capability is particularly promising for applications requiring high-precision astrometry and photometry, such as the detection of compact binary systems and the study of gravitational lensing. The application of the \textit{fiDrizzle-MU} algorithm led to the successful identification and resolution of a new gravitationally lensed quasar candidate, which remained poorly resolved using standard JWST data processing pipelines. Our method clearly distinguished six components of this system and effectively suppressed background noise, highlighting its potential for high-fidelity analysis of complex lensing structures.

Nevertheless, several limitations and areas for further development warrant discussion. First, although {\it fiDrizzle-MU} exhibits faster convergence than {\it fiDrizzle-DC}, its computational complexity scales with the number of dithered frames and the chosen {\tt PSR} parameter. Optimization of computational efficiency, possibly through parallelization strategies or GPU acceleration, will be crucial for processing extremely large datasets, such as those anticipated from upcoming deep-field surveys conducted by CSST-MCI. Second, while positivity constraints serve as an effective regularization prior, additional constraints—such as sparsity priors in a wavelet domain or total variation regularization—could further enhance the algorithm's ability to recover faint and high-frequency features, especially in low signal-to-noise conditions. Integration of adaptive stopping criteria, based on cross-validation or statistical noise modeling, may also improve convergence control and prevent overfitting.


Looking ahead, the {\it fiDrizzle-MU} framework shows substantial promise for high-precision imaging in space-based and ground-based astronomical observations. We plan to apply this method to the extreme deep-field datasets of CSST-MCI, comprising up to 1,600 dithered frames per field, with the goal of enabling precision studies such as dark matter subhalo detection through strong gravitational lens modeling. Further extensions may include the development of hybrid approaches combining {\it fiDrizzle-MU} with PSF deconvolution techniques, to simultaneously address instrumental blurring and dithering-induced aliasing.

In summary, {\it fiDrizzle-MU} represents a robust and efficient image reconstruction algorithm tailored to modern astronomical datasets. Its improvements over prior methods in both convergence behavior and reconstruction accuracy establish it as a valuable tool for precision cosmology and astrophysical imaging. In addition, future work may explore the integration of {\it fiDrizzle-MU} into large-scale data processing pipelines for next-generation surveys such as Euclid or LSST, where robust handling of undersampled dithered data will be critical.

\vspace{0.6cm}

\section*{Acknowledgements}
This work is supported by the Foundation for Distinguished Young Scholars of Jiangsu Province (No. BK20140050), National Key R\&D Program of China No. 2022YFF0503403, the National Natural Science Foundation of China (Nos. 11973070, 11988101, 11333008, 11273061, 11825303, and 11673065), the China Manned Space Project with No. CMS-CSST-2021- A01, CMS-CSST-2021-A03, CMS-CSST-2021-B01 and the Joint Funds of the National Natural Science Foundation of China (No. U1931210). HYS acknowledges the support from the Key Research Program of Frontier Sciences, CAS, Grant No. ZDBS-LY-7013 and Program of Shang- hai Academic/Technology Research Leader. Ran Li acknowledges the science research grants from the China Manned Space Project, CAS Project for Young Scientists in Basic Research (No. YSBR-062), and the support from K.C.Wong Education Foundation. We acknowledge the support from the science research grants from the China Manned Space Project with CMS-CSST-2021-A04, and CMS-CSST-2021-A07. This work is also supported by the GHfund A (202302017475).

\bibliographystyle{raa}
\bibliography{bibtex}

\end{document}